\documentclass[twoside,12pt]{article}
\usepackage{jmlr2e}

\jmlrheading{1}{2019}{--}{5/19}{--/--}{ark007@ufl.edu}{}

\usepackage{amssymb}
\usepackage{amsmath}
\usepackage{graphicx}
\usepackage{comment}
\usepackage{mathrsfs}
\usepackage[colorinlistoftodos]{todonotes}
\usepackage{rotating}
\usepackage{lscape}
\usepackage{epstopdf}
\usepackage{natbib}
\usepackage{color}

\usepackage{enumerate}   
\usepackage{url} 

\ShortHeadings{Nonparametric graphical model for counts}{Roy and Dunson}
\firstpageno{1}

\begin{document}
	
	\title{Nonparametric graphical model for counts}
	
	\author{\name Arkaprava Roy \email ark007@ufl.edu \\
		\addr Department of Biostatistics\\
		University of Florida\\
		Gainesville, FL 32603, USA
		\AND
		\name David B Dunson \email dunson@duke.edu \\
		\addr Department of Statistics\\
		Duke University\\
		Durham, NC 27708-0251, USA}
	
	\editor{}
	
	\maketitle
	
	\bigskip
	
	\begin{abstract}
		Although multivariate count data are routinely collected in many application areas, there is surprisingly little work developing flexible models for characterizing their dependence structure. This is particularly true when interest focuses on inferring the conditional independence graph.  In this article, we propose a new class of pairwise Markov random field-type models for the joint distribution of a multivariate count vector.  By employing a novel type of transformation, we avoid restricting to non-negative dependence structures or inducing other restrictions through truncations. Taking a Bayesian approach to inference, we choose a Dirichlet process prior for the distribution of a random effect to induce great flexibility in the specification.  An efficient Markov chain Monte Carlo (MCMC) algorithm is developed for posterior computation. We prove various theoretical properties, including posterior consistency, and show that our COunt Nonparametric Graphical Analysis (CONGA) approach has good performance relative to competitors in simulation studies.  The methods are motivated by an application to neuron spike count data in mice.
	\end{abstract}
	

	\section{Introduction}
	
	Graphical models provide an appealing framework to characterize dependence in multivariate data $X_i=(X_{i1},\ldots,X_{iP})$ in an intuitive way. This article focuses on undirected graphical models or Markov random fields (MRFs). In this approach, each random variable is assigned as a node of a graph $G$ which is characterized by the pair $(V, E)$. Here $V$ and $E$ denote the set of nodes and set of connected edges of the graph $G$, with $V=\{1,\ldots,P\}$ and $E\subseteq V\times V$. The graph $G$ encodes conditional independence relationships in the data.
	We say $X_{l}$ and $X_{k}$ are conditionally independent if $P(X_{l},X_{k}|X_{-(l,k)})=P(X_{l}|X_{-(l,k)})P(X_{k}|X_{-(l,k)})$, with $X_{-(l,k)}$ denoting all random variables excluding $X_{l}$ and $X_{k}$. Conditional independence between two random variables is equivalent to the absence of an edge between those two corresponding nodes in the graph. Thus the conditional independence of $X_{l}$ and $X_{k}$ would imply that the edge $(k,l)$ is not present i.e. $(k,l)\notin E$. 
	
	Although there is a rich literature on graphical models, most of the focus has been specifically on Gaussian graphical models. For bounded discrete data, Ising \citep{ravikumar2010high,kolar2010estimating} and multinomial graphical models \citep{jalali2011learning} have been studied. However, for unbounded count-valued data, the existing literature is limited. Multivariate count data are routinely collected in genomics, sports, imaging analysis and text mining among many other areas, but most of the focus has been on latent factor and covariance structure models \citep{wedel2003factor,zhou2012beta}. The goal of this article is to address this gap and provide a flexible framework for statistical inference in count graphical models.
	
	Besag first introduced pair-wise graphical models, deemed `auto-models' in his seminal paper on MRFs \citep{besag1974spatial}. To define a joint distribution on a spatial lattice, he started with an exponential family representation of the marginal distributions and then added first order interaction terms. In the special case of count data, he introduced the Poisson auto-model. In this approach, the random variable at the $i$-$th$ location $X_i$ follows a conditional Poisson distribution with mean $\mu_i$, dependent on the neighboring sites. Then $\mu_i$ is given the form $\mu_i=\exp(\alpha_i+\sum_{j}\beta_{ij}X_{j})$. It can be shown that this conditional density model admits a joint density if and only if $\beta_{ij}\leq 0$ for all pairs of $(i, j)$. Hence, only non-negative dependence can be accommodated. Gamma and exponential auto-models also have the same restriction due to non-negativity of the random variables. 
	
	\cite{yang2013poisson} truncated the count support within the Poisson auto-model to allow both positive and negative dependence, effectively treating the data as ordered categorical. \cite{allen2012log} fit the Poisson graphical model locally in a manner that allows both positive and negative dependence, but this approach does not address the problem of global inference on $G$. { {\cite{chiquet2018variational} let $X_{ij}\sim \mbox{Poi}(\mu_j+Z_{ij})$ for $1\leq i\leq n, 1\leq j\leq V$ and $Z_{i}\sim$MVN$(0,\Sigma)$. The graph is inferred through sparse estimation of $\Sigma^{-1}$. \cite{hadiji2015poisson} proposed a non-parametric count model, with the conditional mean of each node an unknown function of the other nodes. \cite{yang2015graphical} defined a pairwise graphical model for count data that only allows negative dependence. \cite{inouye2016generalized,inouye2016square, inouye2017review} models 
			multivariate count data under the assumption that the square root, or more generally the $j$-$th$ root, of the data is in an exponential family. This model allows for positive and negative dependence but under strong distributional assumptions.}}

	In the literature on spatial data analysis, many count-valued spatial processes have been proposed, but much of the focus has been on including spatial random effects instead of an explicit graphical structure. \cite{de2013hierarchical} considered a random field on the mean function of a Poisson model to incorporate spatial dependence. However, conditional independence or dependence structure in the mean does not necessarily represent that of the data. The Poisson-Log normal distribution, introduced by \cite{aitchison1989multivariate}, is popular for analyzing spatial count data \citep{chan1995monte,diggle1998model,chib2001markov,hay2001bayesian}. Here also the graph structure of the mean does not necessarily represent that of the given data. Hence, these models cannot be regarded as graphical models for count data. To study areal data, conditional autoregressive models (CAR) have been proposed \citep{gelfand2003proper,de2012bayesian,wang2013poisson}. Although these models have an MRF-type structure, they assume the graph $G$ is known based on the spatial adjacency structure, while our focus is on inferring unknown $G$. 
	
	Gaussian copula models are popular for multivariate non-normal data \citep{xue2000multivariate,murray2013bayesian}. \cite{mohammadi2017bayesian} developed a computational algorithm to build graphical models based on Gaussian copulas using methods developed by \cite{dobra2011bayesian}. However, it is difficult to model multivariate counts with zero inflated or multimodal marginals using a Gaussian copula. 
	
	Within a semiparametric framework, \cite{liu2009nonparanormal} proposed a nonparanormal graphical model in which an unknown monotone function of the observed data follows a multivariate normal model with unknown mean and precision matrix subject to identifiability restrictions. This model has been popular for continuous data, providing a type of Gaussian copula. For discrete data the model is not directly appropriate, as mapping discrete to continuous data is problematic. To the best of our knowledge, there has been no work on nonparanormal graphical models for counts. In general conditional independence cannot be ensured if the function of the random variable is not continuous. For example if $f$ is not monotone continuous, then conditional independence of $X$ and $Y$ does not ensure conditional independence of $f(X)$ and $f(Y)$.
	
	
	In addition to proposing a flexible graphical model for counts, we aim to develop efficient Bayesian computation algorithms. Bayesian computation for Gaussian graphical models (GGMs) is somewhat well-developed \citep{dobra2011copula,wang2012bayesian,wang2015scaling,mohammadi2015bayesian}. Unfortunately, outside of GGMs, likelihood-based inference is often problematic due to intractable normalizing constants. For example, the normalizing constant in the Ising model is too expensive to compute except for very small $P$. There are approaches related to surrogate likelihood \citep{kolar2008improved} by  relaxation of the log-partition function \citep{banerjee2008model}. \cite{kolar2010estimating} use conditional likelihood. \cite{besag1975statistical} chose a product of conditional likelihoods as a pseudo-likelihood to estimate MRFs. For exponential family random graphs, \cite{van2009framework} compared maximum likelihood and maximum pseudo-likelihood estimates in terms of bias, standard errors, coverage and efficiency. 
	\cite{zhou2009bayesian} numerically compared the estimates from a pseudo-posterior with exact likelihood based estimates and found they are very similar in small samples for Ising and Potts models. Also for pseudo-likelihood based methods asymptotic unbiasedness and consistency have been studied \citep{comets1992consistency,jensen1994asymptotic,mase2000marked,baddeley2000practical}. \cite{pensar2017marginal} showed consistency of marginal pseudo-likelihood for discrete valued MRFs in a Bayesian framework. 
	
	Recently \cite{dobra2018loglinear} used pseudo-likelihood for estimation of their Gaussian copula graphical model. Although pseudo-likelihood is popular in the frequentist domain for count data \citep{inouye2014admixture,ravikumar2010high,yang2013poisson}, its usage is still non standard in Bayesian estimation for count MRFs. This is mainly because calculating conditional densities is expensive for count data due to unbounded support, making posterior computations hard to conduct. We implement an efficient Markov Chain Monte Carlo (MCMC) sampler for our model using pseudo-likelihood and pseudo-posterior formulations. Our approach relies on a provably accurate approximation to the normalizing constant in the conditional likelihood. We also provide a bound for the approximation error due to the evaluation of the normalizing constant numerically. 
	
	
	In Section~\ref{model}, we introduce our novel graphical model. In Section~\ref{theory}, some desirable theoretical results are presented. Then we discuss computational strategies in Section~\ref{comp} and present simulation results in Section~\ref{sim}. We apply our method to neuron spike data in mice in Section~\ref{realana}. We end with some concluding remarks in Section ~\ref{discussion}.
	
	\section{Modeling}
	\label{model}
	
	Before introducing the model, we define some of the Markov properties related to the conditional independence of an undirected graph. A clique of a graph is the set of nodes where every two distinct nodes are adjacent; that is, connected by an edge. Let us define $\mathcal{N}(j)=\{l:(j,l)\in E\}$. For three disjoint sets $A$, $B$ and $C$ of $V$, $A$ is said to be separated from $B$ by $C$ if every path from $A$ to $B$ goes through $C$. A path is an ordered sequence of nodes $i_0, i_1,\ldots, i_m$ such that $(i_{k-1},i_k)\in E$. The joint distribution is locally Markov if $X_j \perp V\setminus\{X_j,N(j)\} | N(j)$. If for three disjoint sets $A, B$ and $C$ of $V$, $X_A$ and $X_B$ are independent given $X_C$ whenever $A$ and $B$ are separated by $C$, the distribution is called globally Markov. The joint density is pair-wise Markov if for any $i,j\in V$ such that $(i,j)\notin E$, $X_i$ and $X_j$ are conditionally independent.
	
	We consider here a pair-wise MRF \citep{wainwright2007high,chen2014selection} which implies the following joint probability mass function (pmf) for the $P$ dimensional random variable $X$,
	\begin{align}
	\mathrm{Pr}(X_1,\ldots,X_P)\propto \exp\bigg\{\sum_{i=1}^Pf(X_i)+\sum_{l=2}^P\sum_{j<l}f(X_j, X_l)\bigg\},\label{pairMRF}
	\end{align}
	where $f(X_i)$ is called a node potential function, $f(X_j, X_l)$ an edge potential function and we have $f(X_j, X_l)=0$ if there is no edge $(j,l)$. Thus this distribution is pair-wise Markov by construction. Then \eqref{pairMRF} satisfies the Hammersley-Clifford theorem \citep{hammersley1971markov}, which states that a probability distribution having a strictly positive density satisfies a Markov property with respect to the undirected graph $G$ if and only if its density can be factorized over the cliques of the graph. Since our pair-wise MRF is pair-wise Markov, we can represent the joint probability mass function as a product of mass functions of the cliques of graph $G$. The existence of such a factorization implies that this distribution has both global and local Markov properties.
	
	Completing a specification of the MRF in~\eqref{pairMRF} requires an explicit choice of the potential functions $f(X_j)$ and $f(X_j, X_l)$. In the Gaussian case, one lets $f(X_j)=-\alpha_jX_j^2$ and $f(X_j,X_l)=-\beta_{jl}X_jX_l$, where $\alpha_{j}$ and $\beta_{jl}$ correspond to the diagonal and off-diagonal elements of the precision matrix $\Sigma^{-1}=\mathrm{cov}(X)^{-1}$. In general, the node potential functions can be chosen to target specific univariate marginal densities. If the marginal distribution is Poisson, the appropriate  node potential function is $f(X_j)=\alpha_{j}X_{j}-\log(X_{j}!)$. One can then choose the edge potential functions to avoid overly restrictive constraints on the dependence structure, such as only allowing non-negative correlations. \cite{yang2013poisson} identify edge potential functions with these properties for count data by truncating the support; for example, to the range observed in the sample. This reduces ability to generalize results, and in practice estimates are sensitive to the truncation level. We propose an alternative construction of the edge potentials that avoids truncation.

	\subsection{Model}
	\label{modelconga}
	We propose the following modified pmf for $P$-dimensional count-valued data $X$, 
	\begin{align}
	\textrm{Pr}(X_{1},\ldots,X_{P})&\propto \exp\bigg(\sum_{j=1}^P [\alpha_{j}X_{j}-\log(X_{j}!)]-\nonumber\sum_{l=2}^P\sum_{j<l}\beta_{jl}F(X_{j})F(X_{l})\bigg),
	\end{align}
	where $F(\cdot)$ is a monotone increasing bounded function with support $[0,\infty)$, $f(X_j)=\alpha_{j}X_{j}-\log(X_{j}!)$ and $f(X_j, X_l)=-\beta_{jl}F(X_{j})F(X_{l})$ using the notation of \eqref{pairMRF}. 
	\begin{lemma}
		Let $F(\cdot)$ be uniformly bounded by $U$, then the normalizing constant, say $A(\alpha,\beta)$, can be bounded as,
		$$\exp\bigg(\sum_{j=1}^P\exp(\alpha_{j}) - U^2\sum_{l=2}^P\sum_{j<l}|\beta_{jl}|\bigg)\leq A(\alpha,\beta)\leq \exp\bigg(\sum_{j=1}^P\exp(\alpha_{j}) +  U^2\sum_{l=2}^P\sum_{j<l}|\beta_{jl}|\bigg).$$
	\end{lemma}
	These bounds can be obtained by elementary calculations. The constant $A(\alpha,\beta)$ is the sum of the above pmf over the support of $X$. The sum reduces to a product of $P$ many exponential series sums after replacing the function $F(\cdot)$ by its maximum. 
	
\begin{figure}[htbp]
		\centering
		\includegraphics[width = 1\textwidth]{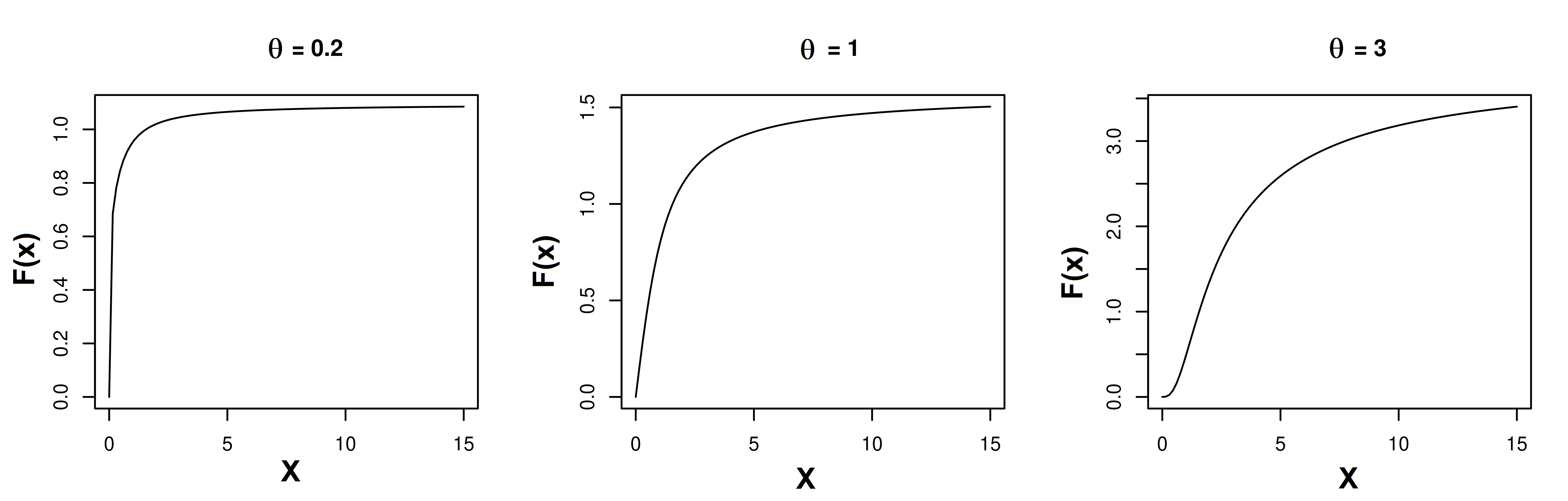}
		\caption{$F(\cdot)=(\tan^{-1})^{\theta}(\cdot)$ for different values of $\theta$. The parameter $\theta$ controls both shape and range of $F(\cdot)$.}
		\label{trans}
	\end{figure}

\begin{figure}[htbp]
		\centering
		\includegraphics[width = 1\textwidth]{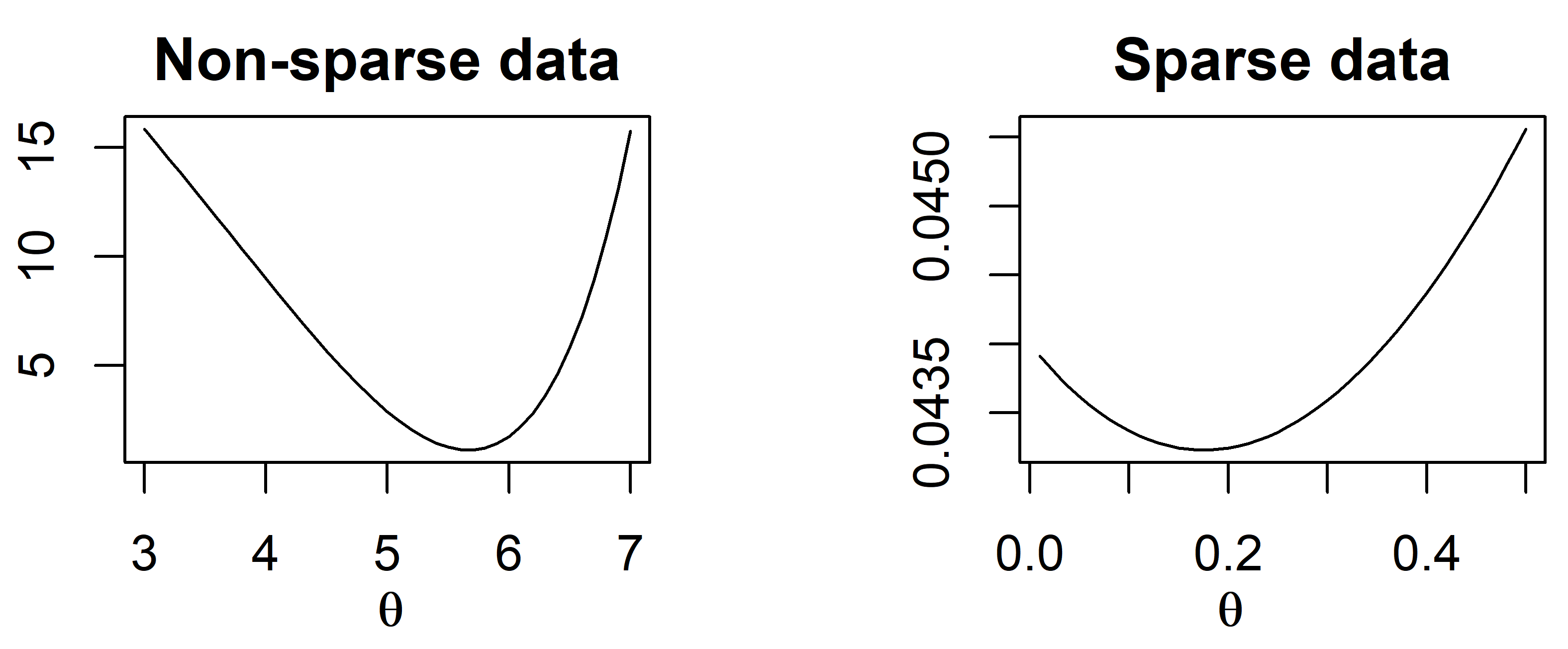}
		\caption{$\|cov(\tan^{-1}(X)^{\theta})-cov(X)\|_F$ for different values of $\theta$. $\|\|_F$ stands for the Frobenius norm.}
		\label{approx}
	\end{figure}

	Thus by modifying the edge potential function in this way using a bounded function of $X$, we can allow unrestricted support for all the parameters, allowing one to estimate both positive and negative dependence. Under the monotonicity restrictions on $F(\cdot)$, inference on the conditional independence structure tends to be robust to the specific form chosen. We let $F(\cdot)=(\tan^{-1}(\cdot))^{\theta}$ for some positive $\theta\in\mathbb{R}^{+}$ to define a flexible class of monotone increasing bounded functions. {  {The exponent $\theta$ provides additional flexibility, including impacting the range of $F(X)$,
			$\big(0,(\frac{\pi}{2})^{\theta}\big)$.  The parameter $\theta$ can be estimated along with the other parameters, including the baseline parameters $\alpha$ controlling the marginal count distributions and the coefficients $\beta_{jl}$ controlling the graphical dependence structure.  For simplicity and interpretability, we propose to estimate $\theta$ to minimize the difference in covariance between $F(X)$ and $X$.}}   Figure~\ref{trans} illustrates how $\theta$ controls the range and shape of $F(\cdot)$. Figure~\ref{approx} shows how the difference between covariances of $F(X)$ and $X$ vary for different values of $\theta$ in sparse and non sparse data cases. In both cases, the difference function has a unique minimizer. { { Although the same strategy could be used to tune the truncation parameter in the \cite{yang2013poisson} approach, issues arise in estimating the support of the data based on a finite sample, as new data may fall outside of the estimated support.  In addition, their approach is less flexible in relying on parametric assumptions, while we use a mixture model for the $\alpha$s to induce a nonparametric structure. }} 
	
	Letting $X_t$ denote the $t^{th}$ independent realization of $X$, for $t=1,\ldots,n$, the pmf is
	\begin{align}
	\textrm{Pr}(X_{t1},\ldots,X_{tP})&\propto \exp\bigg(\sum_{j=1}^P [\alpha_{tj}X_{tj}-\log(X_{tj}!)]-\nonumber\\&\quad\sum_{l=2}^P\sum_{j<l}\beta_{jl}(\tan^{-1}(X_{tj}))^\theta(\tan^{-1}(X_{tl}))^\theta\bigg)\label{NPGMCi},
	\end{align}
	where $\alpha_{tj}$'s are coefficients of different node potential functions and $\beta_{jl}$'s are coefficients of the edge potential functions as before. We vary $\alpha_{tj}$ with $t$ to allow more flexibility in modeling the marginal densities. If $\beta_{jl}=0$, then $X_{tj}$ and $X_{tl}$ are conditionally independent for all $t$. We call our proposed method COunt Nonparametric Graphical Analysis (CONGA). 
	
	Now we reparametrize \eqref{NPGMCi} using $\log(\lambda_{tj})=\alpha_{tj}$ and rewrite the model as,
	\begin{align}
	\textrm{Pr}(X_{t1},\ldots,X_{tP})&\propto \prod_{j=1}^P\frac{\lambda_{tj}^{X_{tj}}}{X_{tj}!}\exp\bigg(-\sum_{l=2}^P\sum_{j<l}\beta_{jl}(\tan^{-1}(X_{tj}))^\theta(\tan^{-1}(X_{tl}))^\theta\bigg)\label{NPGMC}.
	\end{align}
	
	This reparametrizated model is more intuitive to understand. Due to the Poisson type marginal in \eqref{NPGMC}, this model is suitable for data with over-dispersed marginals with respect to the Poisson at each node. Over-dispersion is typical in broad applications. We consider this reparametrized model in the rest of the paper. 
	
	\subsection{Prior structure}
	\label{prior}
	To proceed with Bayesian computation, we put priors on the parameters. We have two sets of parameters in \eqref{NPGMC}, $\beta$ and $\lambda$. For the $\beta_{jl}$ parameters, we choose simple iid Gaussian priors. It is straightforward to consider more elaborate shrinkage or variable selection priors for the $\beta_{jl}$'s, but we find usual Gaussian priors have good performance in small to moderate-dimensional applications 
	
	The parameter $\lambda_{tj}$'s represent random effects; these parameters are not individually identifiable and are given random effects distributions $\lambda_{tj}\sim D_j$. The distribution $D_j$ controls over-dispersion and the shape of the marginal count distribution for the $j^{th}$ node. To allow these marginals to be flexibly determined by the data, we take a Bayesian nonparametric approach using Dirichlet process priors $D_j\sim$DP($M_jD_0$), with $D_0$ a Gamma base measure and $M_j$ a precision parameter, having $M_j\sim $Ga($c,d$) for increased data adaptivity. 
	
	
	\section{Theoretical properties}
	\label{theory}
	We explore some of the theoretical properties of our proposed CONGA method. 
	
	\begin{theorem}
		\label{contheo}
		If we have $\beta_{jl}=0$, then $X_{tj}$ and $X_{tl}$ are conditionally independent for all $t$ under~\eqref{NPGMC}.
	\end{theorem}
	This result is easy to verify by simply calculating the conditional probabilities. The details of the proof are in the Appendix. 
	
	
	
	We study posterior consistency under a fixed $P$ and increasing $n$ regime, assuming the prior of Section~\ref{prior} with prespecified $\theta$. Let $G_j$ be the density on $\alpha_{tj}$, induced by $\lambda_{tj}\sim D_j$. Let the parameter space for $G_j$ be $\mathcal{G}_j$ and that for $\beta$ be $\mathbb{R}^q$, where $q=P(P-1)/2$. Thus the complete parameter space for $\kappa=\{\beta, G_1,\ldots,G_p\}$ is $\Psi=\mathbb{R}^q\times \mathcal{G}_1\times\cdots\times \mathcal{G}_P$. We consider the prior $\tilde{\Gamma}_j$ on $G_j$ and $\chi$ on $\beta$. 
	
	Let $\kappa^0$ be the truth for $\kappa$. We make following assumptions.
	
	{\it Assumptions}
	\begin{enumerate}
		\item For some large $T>0$, let $\mathcal{G}=\{G:G([-T,T])=1\}$. Then ${G}_j^0\in \mathcal{G}$ and ${G}_j^0$ is in the support of $\tilde{\Gamma}_j$.
		\item For some large $C>0$, let $\mathcal{Q}=\{\beta:\|\beta\|_{\infty}<C\}$, where $\|\cdot\|_{\infty}$ stands for the infinity norm. Then $\beta^0\in\mathcal{Q}$ and $\beta^0$ is in the support of $\chi$.
		\item $E(X_{tj})<\infty$ for all pairs of $(t,j)$
	\end{enumerate}
	\begin{theorem}
		\label{consistheo}
		Under the assumptions 1-3, the posterior for $\kappa$ is consistent at $\kappa^0$.
	\end{theorem}
	We show that the truth belongs to the Kullback-Leibler support of the prior. Thus the posterior probability of any neighbourhood around the true p.m.f converges to one in $P_{\kappa_0}^{(n)}$-probability as $n$ goes to $\infty$ as a consequence of \cite{schwartz1965bayes}. Here $P_{\kappa}^{(n)}$ is the distribution of a sample of $n$ observations with parameter $\kappa$. Hence, the posterior is weakly consistent. The posterior is said to be strongly consistent if the posterior probability of any neighbourhood around the true p.m.f convergences to one almost-surely. Support of the data is a countable space. The weak and strong topologies on countable spaces are equivalent by Scheffe's theorem. In particular, weak topology and total variation topology are equivalent for discrete data. Weak consistency implies strong consistency. Thus the posterior for $\kappa$ is also strongly consistent at $\kappa^0$. A detailed proof is in the Appendix. 
	
	Instead of assuming bounded support on the true distribution of random effects, one can also assume it to have sub-Gaussian tails. The posterior consistency result still holds with minor modifications in the current proof. Establishing graph selection consistency of the proposed method is an interesting area of future research when $p$ is growing with $n$ and $\lambda_{tj}$'s are fixed effects. Since, we are interested in a non-parametric graphical model, we do not explore that in this paper.
	
	\section{Computation}
	\label{comp}
	As motivated in Section~\ref{modelconga}, we estimate $\theta$ to minimize the differences in the sample covariance of $X$ and $F(X)$. In particular, the criteria is to minimize $\|cov(\tan^{-1}(X)^{\theta})-cov(X)\|_F$. This is a simple one dimensional optimization problem, which is easily solved numerically. 
	
	To update the other parameters, we use an MCMC algorithm, building on the approach of \cite{roy2018spatial}. We generate proposals for Metropolis-Hastings (MH) using a Gibbs sampler derived under an approximated model. To avoid calculation of the global normalizing constant in the complete likelihood, we consider a pseudo-likelihood corresponding to a product of conditional likelihoods at each node. This requires calculations of $P$ local normalizing constants which is computationally tractable. 
	
	The conditional likelihood at the $j$-$th$ node is,
	\begin{align}
	&P(X_{tj}|X_{t,-j})=\nonumber\\&\quad\frac{\exp\big[\{\log(\lambda_{tj})X_{tj}-\log(X_{tj}!)\}-\sum_{j\neq l}\beta_{jl}\{\tan^{-1}(X_{tj})\}^\theta\{\tan^{-1}(X_{tl})\}^{\theta}\big]}{\sum_{X_{tj}=0}^{\infty}\exp\big[\{\log(\lambda_{tj})X_{tj}-\log(X_{tj}!)\}-\sum_{j\neq l}\beta_{jl}\{\tan^{-1}(X_{tj})\}^\theta\{\tan^{-1}(X_{tl})\}^{\theta}\big]}\label{conlike}
	\end{align}
	The normalizing constant is $$\sum_{X_{tj}=0}^{\infty}\exp\big[\{\log(\lambda_{tj})X_{tj}-\log(X_{tj}!)\}-\sum_{j\neq l}\beta_{jl}\{\tan^{-1}(X_{tj})\}^\theta\{\tan^{-1}(X_{tl})\}^{\theta}\big].$$ We truncate this sum at a sufficiently large value $B$ for the purpose of evaluating the conditional likelihood. The error in this approximation can be bounded by $$\exp(\lambda_{tj})(1-CP(B+1,\lambda_{tj}))\exp\bigg\{-\sum_{j\neq l:\beta_{jl}<0}\beta_{jl}(\pi/2)^\theta(\tan^{-1}(X_{tl}))^{\theta}\bigg\},$$ where $CP(x, l)$ is the cumulative distribution function of the Poisson distribution with mean $l$ evaluated at $x$. The above bound can in turn be bounded by a similar expression with $(\tan^{-1}(X_{tl}))^{\theta}$ replaced by $(\pi/2)^\theta$. One can tune $B$ based on the resulting bound on the approximation error. In our simulation setting, even $B=70$ makes the above bound numerically zero. We use $B=100$ as a default choice for all of our computations.
	
	We update $\lambda_{.j}$ using the MCMC sampling scheme described in Chapter 5 of \cite{ghosal2017fundamentals} for the Dirichlet process mixture prior of $\lambda_{ij}$ based on the above conditional likelihood.  For clarity this algorithm is described below:
	
	\begin{enumerate}[(i)]
		\item Calculate the probability vector $Q_j$ for each $j$ such that $Q_j(k)=\mathrm{Pois}(X_{ij}, \lambda_{kj})$ and $Q_j(i)=M_j\mathrm{Ga}(\lambda_{i, j},a+X_{i, j}, b + 1)$.
		\item Sample an index $l$ from $1:T$ with probability $Q_j/\sum_{k}Q_j(k)$. 
		\item If $l\neq i$, $\lambda_{ij}=\lambda_{lj}$. Otherwise sample a new value as described below.
		\item $M_j$ is sampled from Gamma$(c+U, d-\log(\delta))$, where $U$ is the number of unique elements in $\lambda_{.j}$, $\delta$ is sampled from Beta$(M_j, T)$, and $M_j\sim$Ga($c,d$) a priori.
	\end{enumerate}
	
	When we have to generate a new value for $\lambda_{tj}$ in step (iii), we consider the following scheme.
	\begin{enumerate}[(i)]
		\item Generate a candidate $\lambda_{tj}^c$ from Gamma$(a+X_{tj}, b+1)$.
		\item Adjust the update $\lambda_{tj}^c=\lambda_{tj}^0+K_1(\lambda_{tj}^c-\lambda_{tj}^0)$, where $\lambda_{tj}^0$ is the current value and $K_1<1$ is a tuning parameter, adjusted with respect to the acceptance rate of the resulting Metropolis-Hastings (MH) step.
		\item We use the pseudo-likelihood based on the conditional likelihoods in \eqref{conlike} to calculate the MH acceptance probability. 
	\end{enumerate}
	
	To generate $\beta$, we consider a new likelihood that the standardized $(\tan^{-1}(X_{tl}))^{\theta}$ follows a multivariate  Gaussian distribution with precision matrix $\Omega$ such that $\Omega_{pq}=\Omega_{qp}=\beta_{pq}$ with $p<q$ and $\Omega_{pp}=(Var((\tan^{-1}(X_{tl}))^{\theta})^{-1})_{pp}$. Thus diagonal entries do not change over iterations. We update $\Omega_{l,-l}=\{\Omega_{l,i}:i\neq l\}$ successively. We also define $\Omega_{-l,-l}$ as the submatrix by removing $l$-$th$ row and column. Let $s=(F(x)-\bar{F}(X))^{T}(F(x)-\bar{F}(X))$. Thus $s$ is the $P\times P$ gram matrix of $(\tan^{-1}X)^\theta$, standardized over columns.
	\begin{enumerate}[(i)]
		\item Generate an update for $\Omega_{l,-l}$ using the posterior distribution as in \cite{wang2012bayesian}. Thus a candidate $\Omega_{l,-l}^c$ is generated from MVN$(-Cs_{l,-l}, C)$, where $C=((s_{22}+\gamma)\Omega_{-l,-l}^{-1}+D_l^{-1})^{-1}$, where $D_{l}$ is the prior variance corresponding to $\Omega_{l,-l}$
		\item Adjust the update $\Omega_{l,-l}^c=\Omega_{l,-l}^0+K_2\frac{(\Omega_{l,-l}^c-\Omega_{l,-l}^0)}{\|(\Omega_{l,-l}^c-\Omega_{l,-l}^0)\|_2}$, where $\Omega_{l,-l}^0$ is the current value and $K_2$ is a tuning parameter, adjusted with respect to the acceptance rate of the following MH step. Also $K_2$ should always be less than $\|(\Omega_{l,-l}^c-\Omega_{l,-l}^0)\|_2$. 
		\item Use the pseudo-likelihood based on the conditional likelihoods in \eqref{conlike}, multiplying over $t$ to calculate the MH acceptance probability. $\pi(\theta^0|\theta^c)=\tilde{\pi}(\theta_G)$ and $\pi(\theta^c|\theta^0)=\tilde{\pi}(\theta_G')$, where $\theta_G$ is the original Gibbs update.
	\end{enumerate}
	
	\section{Simulation}
	\label{sim}
	We consider four different techniques for generating multivariate count data. One approach is based on a Gaussian copula type setting. The other three are based on competing methods. We compare the methods based on false positive and false negative proportions. We include an edge in the graph between the $j^{th}$ and $l^{th}$ nodes if the 95\% credible interval for $\beta_{jl}$ does not include zero. There is a decision theoretic proof to justify such an approach in \cite{thulin2014decision}. We compare our method CONGA  with TPGM, SPGM, LPGM, huge, BDgraph and ssgraph. The first three are available in R package {\tt XMRF} and the last two are in R packages {\tt BDgraph} and {\tt ssgraph} respectively. The function huge is from R package {\tt huge} which fits a nonparanormal graphical model. The last two methods fit graphical models using Gaussian copulas and {\tt ssgraph} uses spike and slab priors in estimating the edges.
	
	To simulate data under the first scheme, we follow the steps given below.
	
	\begin{enumerate}[(i)]
		\item Generate $n$ many multivariate normals of length $c$ from MVN$(0_{c}, \Omega^{-1}_{c\times c})$, where $0_c$ is the vector of zeros of length $c$. This produces a matrix $X$ of dimension $n\times c$.
		\item We calculate the matrix $P_{n\times c}$, which is $P_{ij}=\Phi(X_{ij})$, where $\Phi$ is the cumulative density function of the standard normal.
		\item The Poisson random variable $Y_{n\times c}$ is $Y_{ij}=QP(P_{ij},\lambda)$ for a given mean parameter $\lambda$ with QP the quantile function of Poisson($\lambda$). 
	\end{enumerate}
	
	Let $X_{:,l}$ denote the $l$-$th$ column of $X$. In the above data generation setup, $\Omega_{pq}=0$ implies that $Y_{:,p}$ and $Y_{:,q}$ are conditionally independent due to Lemma 3 of \cite{liu2009nonparanormal}. The marginals are allowed to be multimodal at some of the nodes, which is not possible under the other simulation schemes.

	Apart from the above approach, we also generate the data using R package {\tt XMRF} from the models  Sub-Linear Poisson Graphical Model (SPGM), Truncated Poisson graphical Model (TPGM) \citep{yang2013poisson}, and Local Poisson Graphical Model (LPGM) \citep{allen2012log}.
	
	We choose $\nu_3=100$, which is the prior variance of the normal prior of $\beta_{jl}$ for all $j,l$. The choice $\nu_3=100$ makes the prior weakly informative. The parameter $\gamma$ is chosen to be 5 as given in \cite{wang2012bayesian}. For the gamma distribution, we consider $a=b=1$. For the Dirichlet process mixture, we take $c=d=10$. We consider $n=100$ and $P=10, 30,50$. We collect 5000 post burn MCMC samples after burning in 5000 MCMC samples. 
	
	We compare the methods based on two quantities $p_1$ and $p_2$. We define these as $p_1$ = Proportion of falsely connected edges in the estimated graph (false positive) and $p_2$ = Proportion of falsely not connected edges in the estimated graph (false negative). We show the pair $(p_1, p_2)$ in Tables~\ref{com1} to~\ref{com3} for number of nodes 10, 30 and 50. All of these results are based on 50 replications. To evaluate the performance of CONGA, we calculate the proportion of replications where zero is included in the corresponding 95\% credible region, constructed from the MCMC samples for each replication. For the other methods, the results are based on the default regularization as given in the R package {\tt XMRF}. Our proposed method overwhelmingly outperforms the other methods when the data are generated using a Gaussian copula type setting instead of generating from TPGM, SPGM or LPGM. For other cases, our method performs similarly to competing methods when the number of nodes is large. In these cases, the competing methods TPGM, SPGM or LPGM are levering on modeling assumptions that CONGA avoids. CONGA beats BDgraph and ssgraph in almost all the scenarios in terms of false positive proportions. The false negative proportions are comparable. The function `huge' performs similarly to CONGA when the data are generated using TPGM, SPGM and LPGM. But CONGA is better than all other methods when the data are generated using the Gaussian copula type setting. This is likely due to the fact that the other cases correspond to simulating data from one of the competitors models.

	\begin{table}[htbp]
		\caption{Performance of the competing methods against our proposed method with 10 nodes. Top row indicates the method used to estimate and the first column indicates the method used to generate the data. $p_1$ and $p_2$ stand for false positive and false negative proportions.}
		\centering
		\resizebox{0.75\textwidth}{!}{\begin{minipage}{\textwidth}
				\begin{tabular}{|l|l|l|l|l|l|l|l|l|l|l|l|l|l|l|}
					\hline
					& \multicolumn{2}{l|}{CONGA} & \multicolumn{2}{l|}{TPGM} & \multicolumn{2}{l|}{SPGM} & \multicolumn{2}{l|}{LPGM} &\multicolumn{2}{l|}{bdgraph} & \multicolumn{2}{l|}{ssgraph}& \multicolumn{2}{l|}{huge} \\ \hline
					Data generation &&&&&&&&&&&&&&
					\\method&  $p_1$         &  $p_2$         &        $p_1$         &  $p_2$          &        $p_1$         &  $p_2$           &           $p_1$         &  $p_2$&           $p_1$         &  $p_2$&           $p_1$         &  $p_2$&           $p_1$         &  $p_2$            \\ \hline
					Multi-Poisson & 0.08         &  0        &    0.22&0.29&0.21&0.34&0.22&0.29  &0&0.90&0.27&0.07& 0.16& 0.20      \\ \hline
					TPGM & 0.04        & 0.25         &   0.10 &0.02&0.07&0.03&0.10&0.03   &0&0.93&0.30&0.15&0.12& 0.13       \\ \hline
					SPGM& 0.06         & 0.23         &  0.09       &    0.04       &    0.07       & 0.03          &    0.09       &   0.04 &0&0.95&0.28&0.14 &0.12& 0.12        \\ \hline
					LPGM&   0.05       &       0.24   & 0.07   &0.06&0.11&0.07&0.07&0.07  &0&0.92&0.31&0.15 &0.10& 0.09       \\ \hline
				\end{tabular}
		\end{minipage}}
		\label{com1}
	\end{table}
	
	\begin{table}[htbp]
		\centering
		\caption{Performance of the competing methods against our proposed method with 30 nodes. Top row indicates the method used to estimate and the first column indicates the method used to generate the data. $p_1$ and $p_2$ stand for false positive and false negative proportions.}
		\resizebox{0.75\textwidth}{!}{\begin{minipage}{\textwidth}
				\begin{tabular}{|l|l|l|l|l|l|l|l|l|l|l|l|l|l|l|}
					\hline
					& \multicolumn{2}{l|}{CONGA} & \multicolumn{2}{l|}{TPGM} & \multicolumn{2}{l|}{SPGM} & \multicolumn{2}{l|}{LPGM} &\multicolumn{2}{l|}{bdgraph} & \multicolumn{2}{l|}{ssgraph} & \multicolumn{2}{l|}{huge}\\ \hline
					Data generation &&&&&&&&&&&&&&
					\\method&  $p_1$         &  $p_2$         &        $p_1$         &  $p_2$          &        $p_1$         &  $p_2$           &           $p_1$         &  $p_2$&           $p_1$         &  $p_2$&           $p_1$         &  $p_2$ &           $p_1$         &  $p_2$            \\ \hline
					Multi-Poisson & 0        &  0        &    0.08&0.57&0.04&0.76&0.08&0.57 &0.43&0.25&0.42&0.25 &0.13 &0.25      \\ \hline
					TPGM & 0.06         & 0.23         &   0.05 &0.23&0.06&0.23&0.06&0.23 &0.41&0.20&0.37&0.21 &0.09&0.19      \\ \hline
					SPGM& 0.07         & 0.22         &  0.06       &    0.23       &    0.06       & 0.22          &    0.06       &   0.23     &0.40&0.21&0.38&0.21 & 0.08&0.18      \\ \hline
					LPGM&   0.07       &       0.23   & 0.06   &0.22&0.06&0.22&0.06&0.21    &0.39&0.19&0.40&0.22  &0.08 &0.19      \\ \hline
				\end{tabular}
		\end{minipage}}
		\label{com2}
	\end{table}
	
	\begin{table}[htbp]
		\centering
		\caption{Performance of the competing methods against our proposed method with 50 nodes. Top row indicates the method used to estimate and the first column indicates the method used to generate the data. $p_1$ and $p_2$ stand for false positive and false negative proportions.}
		\resizebox{0.75\textwidth}{!}{\begin{minipage}{\textwidth}
				\begin{tabular}{|l|l|l|l|l|l|l|l|l|l|l|l|l|l|l|}
					\hline
					& \multicolumn{2}{l|}{CONGA} & \multicolumn{2}{l|}{TPGM} & \multicolumn{2}{l|}{SPGM} & \multicolumn{2}{l|}{LPGM} &\multicolumn{2}{l|}{bdgraph} & \multicolumn{2}{l|}{ssgraph}& \multicolumn{2}{l|}{huge} \\ \hline
					Data generation &&&&&&&&&&&&&&
					\\method&  $p_1$         &  $p_2$         &        $p_1$         &  $p_2$          &        $p_1$         &  $p_2$           &           $p_1$         &  $p_2$&           $p_1$         &  $p_2$&           $p_1$         &  $p_2$  &$p_1$         &  $p_2$          \\ \hline
					Multi-Poisson & 0         &  0        &    0.01&0.88&0.02&0.76&0.02&0.75 &0.46&0.22&0.44&0.25 &0.15 &0.26      \\ \hline
					TPGM & 0.11         & 0.23         &   0.03 &0.29&0.03&0.33&0.03&0.33   &0.42&0.23&0.43&0.25&0.07&0.21       \\ \hline
					SPGM& 0.11         & 0.25         &  0.03       &    0.33       &    0.03       & 0.31        &    0.03       &   0.33   &0.43&0.21&0.41&0.26    &0.08 & 0.22       \\ \hline
					LPGM&   0.12       &       0.23   & 0.03   &0.32&0.03&0.34&0.03&0.31 &0.43&0.23&0.44&0.26  &0.08&0.21          \\ \hline
				\end{tabular}
		\end{minipage}}
		\label{com3}
	\end{table}
	
	\section{Neuron spike count application}
	\label{realana}
\begin{figure}[htbp]
		\centering
		\includegraphics[width = 1\textwidth]{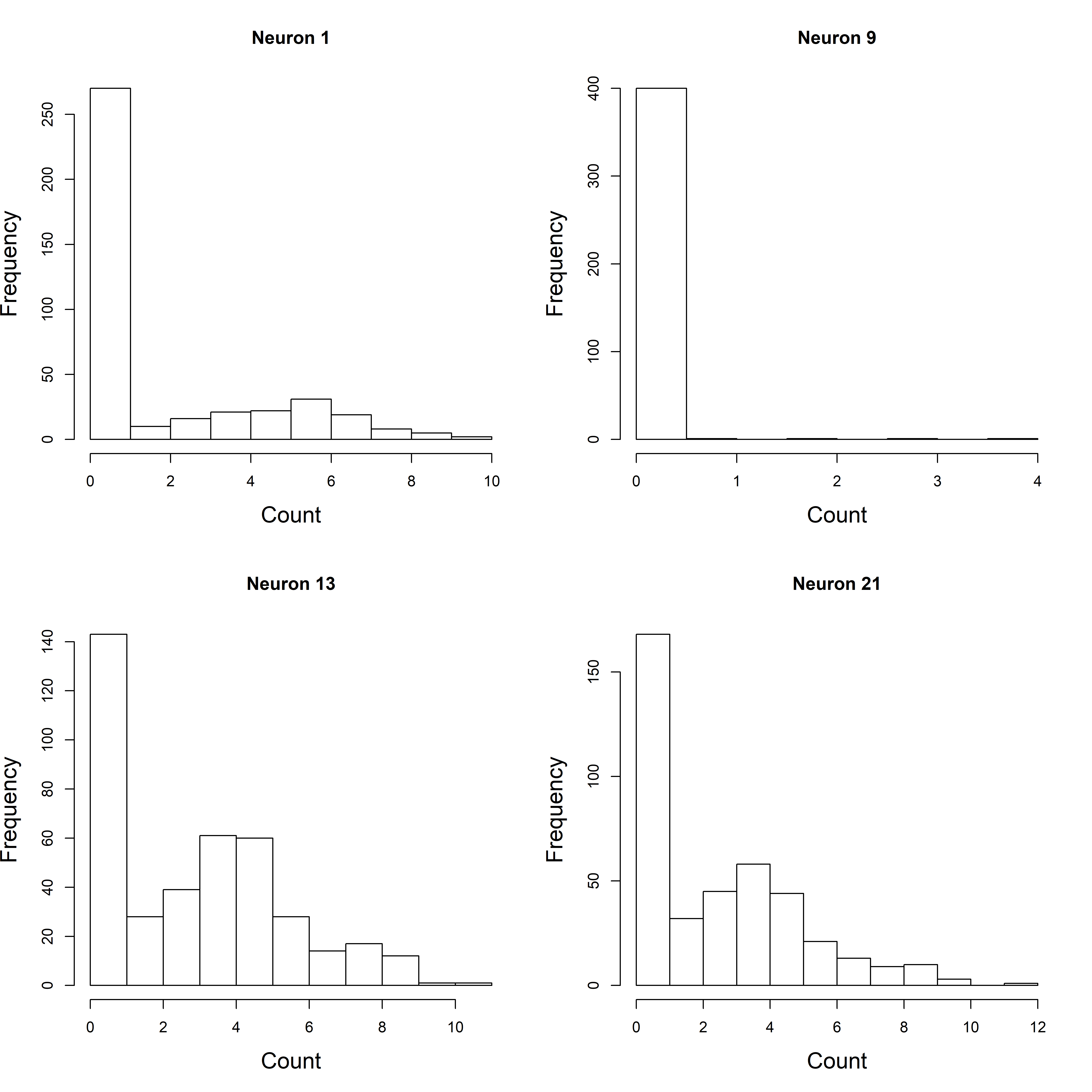}
		\caption{Marginal densities of spike count of the four selected neurons under the influence of stimuli 0.}
		\label{real}
	\end{figure}

	The dataset records neuron spike counts in mice across 37 neurons in the brain under the influence of three different external stimuli, 2-D sinusoids with vertical gradient, horizontal gradient, and the sum. These neurons are from the same depth of the visual cortex of a mouse. The data are collected for around 400 time points. In Figure~\ref{real}, we plot the marginal densities of the spike counts of four neurons under the influence of stimuli 0. We see that there are many variations in the marginal densities, and the densities are multi-modal for some of the cases. Marginally at each node, we also have that the variance is more than the corresponding mean for each of the three stimuli.
	
	\subsection{Estimation}
	We apply exactly the same computational approach as used in the simulation studies. To additionally obtain a summary of the weight of evidence of an edge between nodes $j$ and $l$, we calculate $S_{jl}=\big(|0.5-P(\beta_{jl}>0)|\big)/0.5$, with $P(\beta_{jl}>0)$ the posterior probability estimated from the MCMC samples. We plot the estimated graph with edge thickness proportional to the values of $S_{jl}$. Thus thicker edges suggest greater evidence of an edge in Figures~\ref{network1} to~\ref{network3}. To test for similarity in the graph across stimuli, we estimate 95\% credible regions for $\Delta_{jl}^{s,s'}=\beta_{jl}^s-\beta_{jl}^{s'}$, denoting the difference in the $(j,l)$ edge weight parameter under stimuli $s$ and $s'$, respectively. We flag those edges $(j,l)$ having 95\% credible intervals for $\Delta_{jl}^{s,s'}$ not including zero as significantly different across stimuli.
	
	\subsection{Inference}
\begin{figure}[htbp]
		\centering
		\includegraphics[width = 0.5\textwidth]{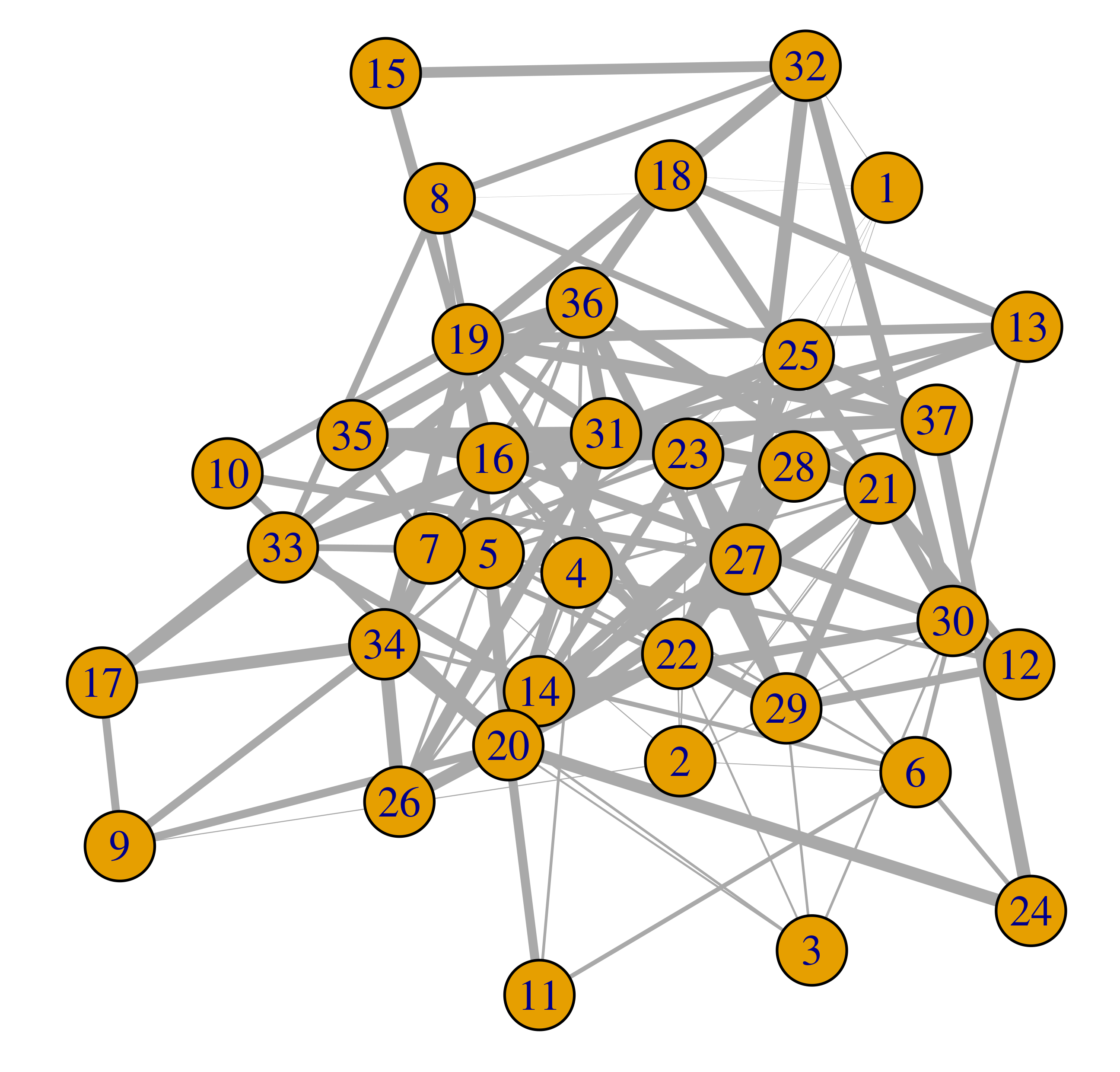}
		\caption{Estimated weighted network under the influence of stimuli 0. The edge width is proportional to the degree of significance.}
		\label{network1}
	\end{figure}

\begin{figure}[htbp]
		\centering
		\includegraphics[width = 0.4\textwidth]{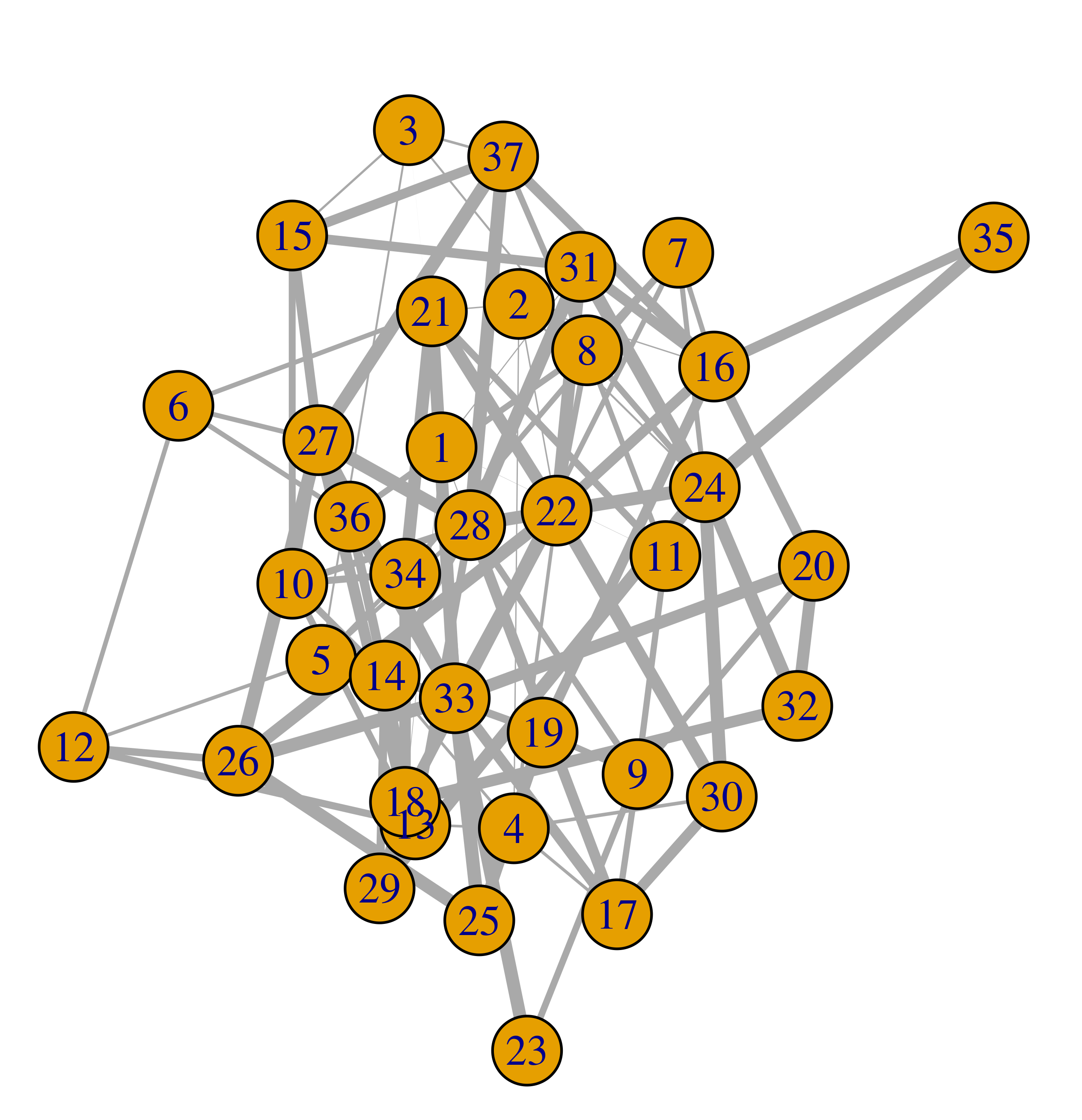}
		\caption{Estimated weighted network under the influence of stimuli 2. The edge width is proportional to the degree of significance.}
		\label{network3}
	\end{figure}

	We find that there are 129, 199 and 110 connected edges respectively for stimuli 0, 1 and 2. Among these edges, 38 are common for stimulus 0 and 1. The number is 15 for stimulus 0 and 2, and 28 for stimulus 1 and 2. There are 6 edges that are common for all of the stimuli. These are (13,16), (8,27), (5,8), (33,35), (3,4) and (9, 14). Each node has at least one edge with another node. We plot the estimated network in Figures~\ref{network1} to~\ref{network3}. We calculate the number of connected edges for each node and list the 5 most connected nodes in Table~\ref{table1}. We also list the most significant 10 edges for each stimulus in Table~\ref{sigedge}. We find that the node 27 is present in all of them. This node seems to have significant interconnections with other nodes for all of the stimuli. We also test the similarity in the estimated weighted network across stimulus. Here we find 82.13\% similarity between the estimated weighted networks under the influence of stimulus 0 and 1. It is 84.98\% for the pair 0 and 2. For 1 and 2, it is 79.43\%. Stimulus 0 is a combination of stimuli 1 and 2. This could be the reason that the estimated graph under influence of stimulus 0 has the highest similarity with the other estimated graphs. 
	
	\begin{figure}[htbp]
		\centering
		\includegraphics[width = 0.4\textwidth]{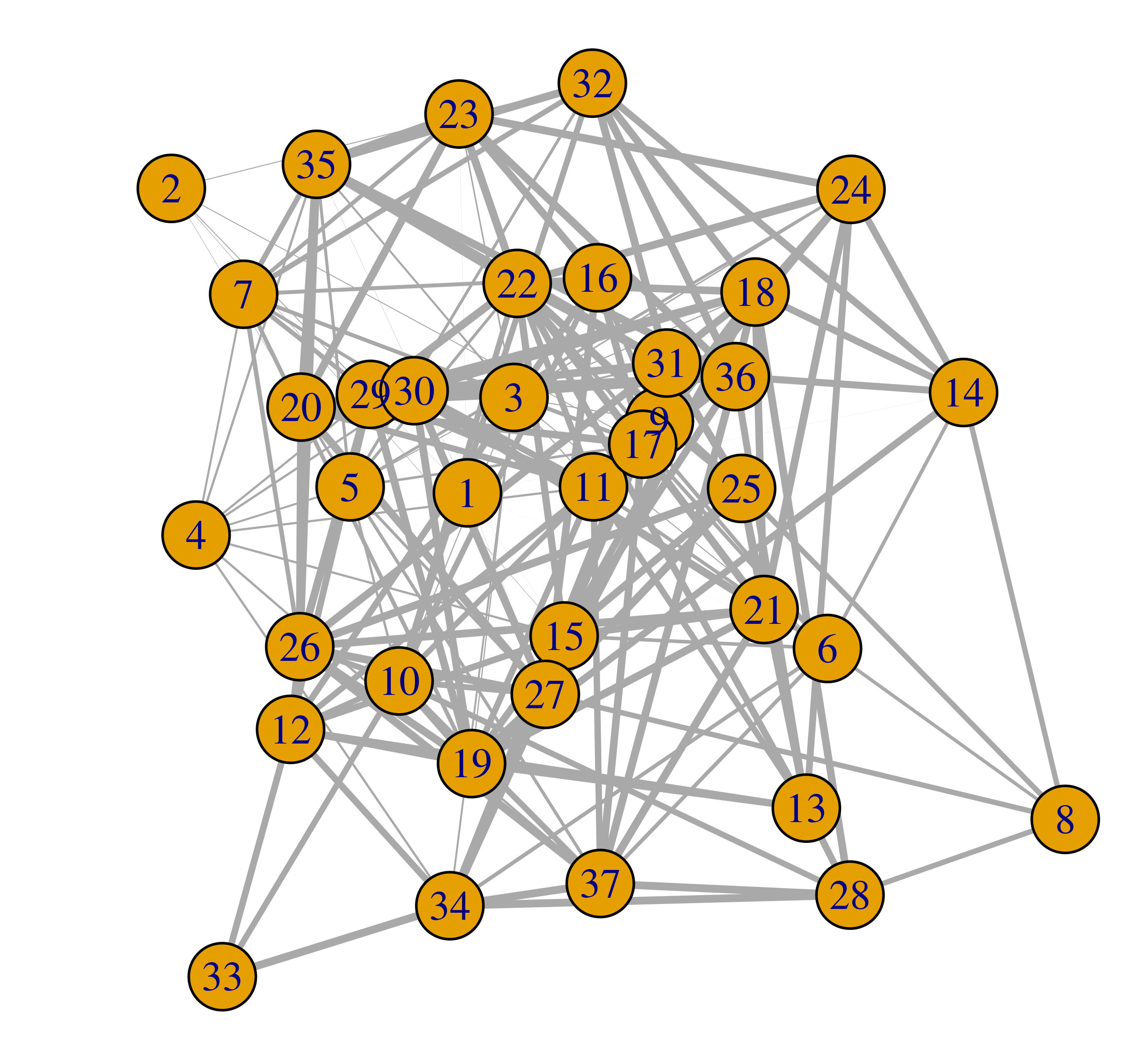}
		\caption{Estimated weighted network under the influence of stimuli 1. The edge width is proportional to the degree of significance.}
	\end{figure}
	
	\begin{table}[htbp]
		\caption{Top 5 nodes with maximum number of connected edges under the influence of stimuli 0, 1 and 2 are listed below.}
		\centering
		\begin{tabular}{|l|l|l|l|l|l|}
			\hline
			\multicolumn{2}{|l|}{Stimuli 0} & \multicolumn{2}{l|}{Stimuli 1} & \multicolumn{2}{l|}{Stimuli 2} \\ \hline
			Node & Number of  &Node & Number of  &Node & Number of \\ 
			number  &connected edges&number &connected edges&number &connected edges\\
			\hline
			37 & 12 & 27 & 16 & 32 & 11 \\ 
			6 & 11 & 3 & 15 & 23 & 10 \\ 
			9 & 11 & 5 & 14 & 3 & 9 \\ 
			25 & 11 & 8 & 14 & 18 & 9 \\ 
			27 & 11 & 23 & 14 & 27 & 9 \\ 
			\hline
		\end{tabular}
		\label{table1}
	\end{table}
	
	\begin{table}[htbp]
		\caption{Top 10 most significant edges under the influence of stimulus 0, 1 and 2 with 1 as the estimated measure of significance are listed below.}
		\centering
		\begin{tabular}{|l|l|l|l|l|l|}
			\hline
			\multicolumn{2}{|l|}{Stimuli 0} & \multicolumn{2}{l|}{Stimuli 1} & \multicolumn{2}{l|}{Stimuli 2} \\ \hline
			Neuron 1 & Neuron 2&Neuron 1 & Neuron 2&Neuron 1 & Neuron 2 \\ 
			\hline
			24 &  35 &  24 &  28 &  14 &  30 \\ 
			26 &  30 &  24 &  30 &  16 &  35 \\ 
			26 &  37 &  24 &  35 &  21 &  35 \\ 
			28 &  37 &  24 &  37 &  21 &  36 \\ 
			29 &  33 &  26 &  28 &  24 &  28 \\ 
			30 &  32 &  26 &  31 &  24 &  29 \\ 
			30 &  35 &  28 &  37 &  24 &  37 \\ 
			31 &  33 &  34 &  37 &  25 &  26 \\ 
			35 &  36 &  35 &  36 &  26 &  36 \\ 
			35 &  37 &  36 &  37 &  31 &  36 \\ 
			\hline
		\end{tabular}
		\label{sigedge}
	\end{table}

	\section{Discussion}
	\label{discussion}
	Our count nonparametric graphical analysis (CONGA) method is useful for multivariate count data, and represents a starting point for more elaborate models and other research directions. One important direction is to time series data. In this respect, a simple extension is to define an autoregressive process for the baseline parameters $\alpha_{tj}$, inducing correlation in $\alpha_{t-1,j}$ and $\alpha_{tj}$, while leaving the graph as fixed in time. A more elaborate extension would instead allow the graph to evolve dynamically by replacing the $\beta_{jl}$ parameters with $\beta_{tjl}$, again defining an appropriate autoregressive process.
	
	{ {In this paper, we proposed to tune $\theta$ by minimizing the difference $\|cov((\tan^{-1}(X))^{\theta})-cov(X)\|_{F}$. However, we could have easily placed a prior on $\theta$ and updated it within our posterior sampling algorithm.  As the gradient of the pseudo-likelihood with respect to $\theta$ is easy to compute, it is possible to develop efficient gradient-based updating algorithms.  When $\lambda_{tj}$'s are fixed effects, an interesting area of research is to establish graph selection consistency.  Such theory would likely 
			give us more insight regarding the role of $\theta$. Graph selection is expected to suffer both for too small and too large $\theta$. }} 
	
	An additional interesting direction is flexible graphical modeling of continuous positive-valued multivariate data. Such a modification is conceptually straightforward by changing the term $\log(X_{tj}!)$ to the corresponding term in the gamma distribution. All the required functions to fit the CONGA algorithm along with a supplementary R code with an example usage are provided at {\url  {https://github.com/royarkaprava/CONGA}}.
	
	\acks{This research was partially supported by
		grant R01-ES027498-01A1 from the National Institute of Environmental Health Sciences (NIEHS) of the National
		Institutes of Health (NIH) and grant R01-MH118927 from the National Institute of Mental Health (NIMH) of the NIH.}

	\newpage
	
	\appendix
	\section*{Appendix}
	\label{appen}
	\subsection*{Proof of Theorem~\ref{contheo}}
	
	The conditional probability is given by,
	\begin{align*}
	&P(X_{tj}, X_{tl}|X_{t,-(j,l)})=\nonumber\\&\quad\frac{\exp\big(\sum_{h\in (j,l)}(\alpha_{th}X_{th}-log(X_{th}!)-\sum_{g\neq h}\beta_{gh}(\tan^{-1}(X_{tg}))^\theta(\tan^{-1}(X_{th}))^{\theta}\big)}{\sum_{X_{tj}=0}^{\infty}\sum_{X_{tl}=0}^{\infty}\exp\big(\sum_{h\in (j,l)}(\alpha_{th}X_{th}-log(X_{th}!)-\sum_{g\neq h}\beta_{gh}(\tan^{-1}(X_{tg}))^\theta(\tan^{-1}(X_{th}))^{\theta}\big)},
	\end{align*}
	where $X_{t,-(j,l)}=\{X_{ti}:i\neq (j,l)\}$ and $\log(\lambda_{th})=\alpha_{th}$. Since $\beta_{jl}=0$, we can break the exponentiated terms into two such that $X_{tj}$ and $X_{jl}$ are separated out. That would immediately give us, $P(X_{tj}, X_{tl}|X_{t,-(j,l)}) = P(X_{tj}|X_{t,-(j,l)})P(X_{tl}|X_{t,-(j,l)})$.

	\subsection*{Proof of Theorem~\ref{consistheo}}
	For $q, q^* \in$ the space of probability measure $ \mathcal{P}$, let the Kullback-Leibler divergences be given by 
	$$
	K(q^*, q) = \int q^*\log{\frac{q^*}{q}} \qquad  V(q^*, q) = \int q^*\log^2{\frac{q^*}{q}}.
	$$
	Let us denote $p_{i,\alpha,\beta}(X_i)$ as the probability distribution of the data given below,
	$$
	\frac{1}{A(\alpha_{i},\beta)}\exp\bigg(\sum_{j=1}^P [\alpha_{ij}X_{ij}-\log(X_{ij}!)]+\sum_{l=2}^P\sum_{j<l}\beta_{ijl}(\tan^{-1}(X_{ij}))^\theta(\tan^{-1}(X_{il}))^\theta\bigg),
	$$
	where $A(\alpha_{i},\beta)$ is the normalizing constant and $\alpha_i=\{\alpha_{i1},\ldots,\alpha_{iP}\}, \beta=\{\beta_{jl}:1\leq j<l\leq P\}$. Let $E(X_{ij})=Q$. We have,
	
	$$\frac{\partial\log(A(\alpha_{i},\beta))}{\partial \alpha_{ij}}=\frac{A(\alpha_{ij},\beta)}{A(\alpha_{i},\beta))}E(X_{ij})\leq Q, \quad \mathrm{ as },\quad \frac{A(\alpha_{ij},\beta)}{A(\alpha_{i},\beta))}\leq 1,$$
	
	and
	
	$$\frac{\partial (A(\alpha_{i},\beta))}{\partial \beta_{jl}}\leq (\frac{\pi}{2})^{2\theta}(A(\alpha_{i},\beta))$$
	
	Thus we have, $\frac{\partial\log(A(\alpha_{i},\beta))}{\partial \alpha_{ik}}\leq Q$ for all $1\leq k\leq P$ and $\frac{\partial\log(A(\alpha_{i},\beta))}{\partial \beta_{jl}}\leq (\frac{\pi}{2})^{2\theta}$.
	
	This implies,
	$$
	-\sum_{j=1}^Pv_{ij}\leq\log\frac{p_{i,\kappa^0}(X_i)}{p_{i,\kappa}(X_i)}\leq\sum_{j=1}^Pv_{ij},
	$$
	where $v_{tj}=(TQX_{tj}+C\sum_{l=2}^P\sum_{j<l}(\tan^{-1}(X_{tj}))+(\frac{\pi}{2})^{2\theta}(T+qC)$. We have $E(v_{tj})<\infty$ due to the last assumption. From the dominated convergence theorem as $n\rightarrow\infty$, we have $\kappa$ converges to $\kappa^0$. Thus Kullback-Leibler divergences go to zero. 
	
	Thus the posterior is weakly consistent. The weak and strong topologies on countable spaces are equivalent by Scheffe's theorem. Thus the posterior for $\kappa$ is also strongly consistent at $\kappa^0$.
	
	\bibliographystyle{bibstyle}
	\bibliography{main}

@article{chib2001markov,
  title={Markov chain Monte Carlo analysis of correlated count data},
  author={Chib, Siddhartha and Winkelmann, Rainer},
  journal={Journal of Business \& Economic Statistics},
  volume={19},
  number={4},
  pages={428--435},
  year={2001},
  publisher={Taylor \& Francis}
}

@book{ghosal2017fundamentals,
  title={Fundamentals of nonparametric Bayesian inference},
  author={Ghosal, Subhashis and Van der Vaart, Aad},
  volume={44},
  year={2017},
  publisher={Cambridge University Press}
}

@article{van2009framework,
  title={A framework for the comparison of maximum pseudo-likelihood and maximum likelihood estimation of exponential family random graph models},
  author={Van Duijn, Marijtje AJ and Gile, Krista J and Handcock, Mark S},
  journal={Social Networks},
  volume={31},
  number={1},
  pages={52--62},
  year={2009},
  publisher={Elsevier}
}

@article{schwartz1965bayes,
  title={On bayes procedures},
  author={Schwartz, Lorraine},
  journal={Zeitschrift f{\"u}r Wahrscheinlichkeitstheorie und verwandte Gebiete},
  volume={4},
  number={1},
  pages={10--26},
  year={1965},
  publisher={Springer}
}

@article{pensar2017marginal,
  title={Marginal pseudo-likelihood learning of discrete Markov network structures},
  author={Pensar, Johan and Nyman, Henrik and Niiranen, Juha and Corander, Jukka and others},
  journal={Bayesian analysis},
  volume={12},
  number={4},
  pages={1195--1215},
  year={2017},
  publisher={International Society for Bayesian Analysis}
}

@article{xue2000multivariate,
  title={Multivariate dispersion models generated from Gaussian copula},
  author={Xue-Kun Song, Peter},
  journal={Scandinavian Journal of Statistics},
  volume={27},
  number={2},
  pages={305--320},
  year={2000},
  publisher={Wiley Online Library}
}

@article{murray2013bayesian,
  title={Bayesian Gaussian copula factor models for mixed data},
  author={Murray, Jared S and Dunson, David B and Carin, Lawrence and Lucas, Joseph E},
  journal={Journal of the American Statistical Association},
  volume={108},
  number={502},
  pages={656--665},
  year={2013},
  publisher={Taylor \& Francis Group}
}

@article{dobra2018loglinear,
  title={Loglinear model selection and human mobility},
  author={Dobra, Adrian and Mohammadi, Reza and others},
  journal={The Annals of Applied Statistics},
  volume={12},
  number={2},
  pages={815--845},
  year={2018},
  publisher={Institute of Mathematical Statistics}
}

@article{mohammadi2017bayesian,
  title={Bayesian modelling of Dupuytren disease by using Gaussian copula graphical models},
  author={Mohammadi, Abdolreza and Abegaz, Fentaw and van den Heuvel, Edwin and Wit, Ernst C},
  journal={Journal of the Royal Statistical Society: Series C (Applied Statistics)},
  volume={66},
  number={3},
  pages={629--645},
  year={2017},
  publisher={Wiley Online Library}
}

@article{kolar2008improved,
  title={Improved estimation of high-dimensional Ising models},
  author={Kolar, Mladen and Xing, Eric P},
  journal={arXiv preprint arXiv:0811.1239},
  year={2008}
}

@article{banerjee2008model,
  title={Model selection through sparse maximum likelihood estimation for multivariate aussian or binary data},
  author={Banerjee, Onureena and Ghaoui, Laurent El and d’Aspremont, Alexandre},
  journal={Journal of Machine Learning Research},
  volume={9},
  number={Mar},
  pages={485--516},
  year={2008}
}

@article{de2012bayesian,
  title={Bayesian analysis of conditional autoregressive models},
  author={De Oliveira, Victor},
  journal={Annals of the Institute of Statistical Mathematics},
  volume={64},
  number={1},
  pages={107--133},
  year={2012},
  publisher={Springer}
}

@article{wang2013poisson,
  title={A Poisson-lognormal conditional-autoregressive model for multivariate spatial analysis of pedestrian crash counts across neighborhoods},
  author={Wang, Yiyi and Kockelman, Kara M},
  journal={Accident Analysis \& Prevention},
  volume={60},
  pages={71--84},
  year={2013},
  publisher={Elsevier}
}

@article{gelfand2003proper,
  title={Proper multivariate conditional autoregressive models for spatial data analysis},
  author={Gelfand, Alan E and Vounatsou, Penelope},
  journal={Biostatistics},
  volume={4},
  number={1},
  pages={11--15},
  year={2003},
  publisher={Oxford University Press}
}

@article{wedel2003factor,
  title={Factor models for multivariate count data},
  author={Wedel, Michel and B{\"o}ckenholt, Ulf and Kamakura, Wagner A},
  journal={Journal of Multivariate Analysis},
  volume={87},
  number={2},
  pages={356--369},
  year={2003},
  publisher={Elsevier}
}

@article{zhou2012beta,
  title={Beta-negative binomial process and Poisson factor analysis},
  author={Zhou, Mingyuan and Hannah, Lauren A and Dunson, David B and Carin, Lawrence},
  journal={In AISTATS},
pages={1462--1471},
  year={2012}
}

@inproceedings{jalali2011learning,
  title={On learning discrete graphical models using group-sparse regularization},
  author={Jalali, Ali and Ravikumar, Pradeep and Vasuki, Vishvas and Sanghavi, Sujay},
  booktitle={Proceedings of the Fourteenth International Conference on Artificial Intelligence and Statistics},
  pages={378--387},
  year={2011}
}

@article{besag1975statistical,
  title={Statistical analysis of non-lattice data},
  author={Besag, Julian},
  journal={The Statistician},
  pages={179--195},
  year={1975},
  publisher={JSTOR}
}

@article{comets1992consistency,
  title={On consistency of a class of estimators for exponential families of Markov random fields on the lattice},
  author={Comets, Francis},
  journal={The Annals of Statistics},
  pages={455--468},
  year={1992},
  publisher={JSTOR}
}

@article{jensen1994asymptotic,
  title={On asymptotic normality of pseudo likelihood estimates for pairwise interaction processes},
  author={Jensen, Jens Ledet and K{\"u}nsch, Hans R},
  journal={Annals of the Institute of Statistical Mathematics},
  volume={46},
  number={3},
  pages={475--486},
  year={1994},
  publisher={Springer}
}

@article{mase2000marked,
  title={Marked Gibbs processes and asymptotic normality of maximum pseudo-likelihood estimators},
  author={Mase, Shigeru},
  journal={Mathematische Nachrichten},
  volume={209},
  number={1},
  pages={151--169},
  year={2000},
  publisher={Wiley Online Library}
}

@article{baddeley2000practical,
  title={Practical Maximum Pseudolikelihood for Spatial Point Patterns: (with Discussion)},
  author={Baddeley, Adrian and Turner, Rolf},
  journal={Australian \& New Zealand Journal of Statistics},
  volume={42},
  number={3},
  pages={283--322},
  year={2000},
  publisher={Wiley Online Library}
}

@inproceedings{inouye2014admixture,
  title={Admixture of Poisson MRFs: A topic model with word dependencies},
  author={Inouye, David and Ravikumar, Pradeep and Dhillon, Inderjit},
  booktitle={International Conference on Machine Learning},
  pages={683--691},
  year={2014}
}

@techreport{zhou2009bayesian,
  title={Bayesian parameter estimation in Ising and Potts models: A comparative study with applications to protein modeling},
  author={Zhou, Xiang and Schmidler, Scott C},
  year={2009},
  institution={Duke University}
}

@article{ravikumar2010high,
  title={High-dimensional Ising model selection using $\ell$1-regularized logistic regression},
  author={Ravikumar, Pradeep and Wainwright, Martin J and Lafferty, John D and others},
  journal={The Annals of Statistics},
  volume={38},
  number={3},
  pages={1287--1319},
  year={2010},
  publisher={Institute of Mathematical Statistics}
}

@article{kolar2010estimating,
  title={Estimating time-varying networks},
  author={Kolar, Mladen and Song, Le and Ahmed, Amr and Xing, Eric P and others},
  journal={The Annals of Applied Statistics},
  volume={4},
  number={1},
  pages={94--123},
  year={2010},
  publisher={Institute of Mathematical Statistics}
}

@article{hammersley1971markov,
  title={Markov fields on finite graphs and lattices},
  author={Hammersley, John M and Clifford, Peter},
  year={1971}
}

@inproceedings{wainwright2007high,
  title={High-Dimensional Graphical Model Selection Using $\ell_1$ Regularized Logistic Regression},
  author={Wainwright, Martin J and Lafferty, John D and Ravikumar, Pradeep K},
  booktitle={Advances in neural information processing systems},
  pages={1465--1472},
  year={2007}
}

@article{chen2014selection,
  title={Selection and estimation for mixed graphical models},
  author={Chen, Shizhe and Witten, Daniela M and Shojaie, Ali},
  journal={Biometrika},
  volume={102},
  number={1},
  pages={47--64},
  year={2014},
  publisher={Oxford University Press}
}

@article{thulin2014decision,
  title={Decision-theoretic justifications for Bayesian hypothesis testing using credible sets},
  author={Thulin, M{\aa}ns},
  journal={Journal of Statistical Planning and Inference},
  volume={146},
  pages={133--138},
  year={2014},
  publisher={Elsevier}
}

@article{hay2001bayesian,
  title={Bayesian analysis of a time series of counts with covariates: an application to the control of an infectious disease},
  author={Hay, John L and Pettitt, Anthony N},
  journal={Biostatistics},
  volume={2},
  number={4},
  pages={433--444},
  year={2001},
  publisher={Oxford University Press}
}

@article{chan1995monte,
  title={Monte Carlo EM estimation for time series models involving counts},
  author={Chan, KS and Ledolter, Johannes},
  journal={Journal of the American Statistical Association},
  volume={90},
  number={429},
  pages={242--252},
  year={1995},
  publisher={Taylor \& Francis}
}

@article{diggle1998model,
  title={Model-based geostatistics},
  author={Diggle, Peter J and Tawn, JA and Moyeed, RA},
  journal={Journal of the Royal Statistical Society: Series C (Applied Statistics)},
  volume={47},
  number={3},
  pages={299--350},
  year={1998},
  publisher={Wiley Online Library}
}

@article{aitchison1989multivariate,
  title={The multivariate Poisson-log normal distribution},
  author={Aitchison, John and Ho, CH},
  journal={Biometrika},
  volume={76},
  number={4},
  pages={643--653},
  year={1989},
  publisher={Oxford University Press}
}

@article{de2013hierarchical,
  title={Hierarchical Poisson models for spatial count data},
  author={De Oliveira, Victor},
  journal={Journal of Multivariate Analysis},
  volume={122},
  pages={393--408},
  year={2013},
  publisher={Elsevier}
}

@article{dobra2011copula,
  title={Copula Gaussian graphical models and their application to modeling functional disability data},
  author={Dobra, Adrian and Lenkoski, Alex and others},
  journal={The Annals of Applied Statistics},
  volume={5},
  number={2A},
  pages={969--993},
  year={2011},
  publisher={Institute of Mathematical Statistics}
}

@article{besag1974spatial,
  title={Spatial interaction and the statistical analysis of lattice systems},
  author={Besag, Julian},
  journal={Journal of the Royal Statistical Society. Series B (Methodological)},
  pages={192--236},
  year={1974},
  publisher={JSTOR}
}

@article{dobra2011bayesian,
  title={Bayesian inference for general Gaussian graphical models with application to multivariate lattice data},
  author={Dobra, Adrian and Lenkoski, Alex and Rodriguez, Abel},
  journal={Journal of the American Statistical Association},
  volume={106},
  number={496},
  pages={1418--1433},
  year={2011},
  publisher={Taylor \& Francis}
}

@article{roy2018spatial,
  title={Spatial shrinkage via the product independent Gaussian process prior},
  author={Roy, Arkaprava and Reich, Brian J and Guinness, Joseph and Shinohara, Russell T and Staicu, Ana-Maria},
  journal={arXiv preprint arXiv:1805.03240},
  year={2018}
}

@article{mohammadi2015bayesian,
  title={Bayesian structure learning in sparse Gaussian graphical models},
  author={Mohammadi, Abdolreza and Wit, Ernst C and others},
  journal={Bayesian Analysis},
  volume={10},
  number={1},
  pages={109--138},
  year={2015},
  publisher={International Society for Bayesian Analysis}
}

@article{wang2015scaling,
  title={Scaling it up: Stochastic search structure learning in graphical models},
  author={Wang, Hao},
  journal={Bayesian Analysis},
  volume={10},
  number={2},
  pages={351--377},
  year={2015},
  publisher={International Society for Bayesian Analysis}
}

@inproceedings{allen2012log,
  title={A log-linear graphical model for inferring genetic networks from high-throughput sequencing data},
  author={Allen, Genevera I and Liu, Zhandong},
  booktitle={Bioinformatics and Biomedicine (BIBM), 2012 IEEE International Conference on},
  pages={1--6},
  year={2012},
  organization={IEEE}
}

@article{liu2009nonparanormal,
  title={The nonparanormal: Semiparametric estimation of high dimensional undirected graphs},
  author={Liu, Han and Lafferty, John and Wasserman, Larry},
  journal={Journal of Machine Learning Research},
  volume={10},
  number={Oct},
  pages={2295--2328},
  year={2009}
}

@article{wang2012bayesian,
  title={Bayesian graphical lasso models and efficient posterior computation},
  author={Wang, Hao},
  journal={Bayesian Analysis},
  volume={7},
  number={4},
  pages={867--886},
  year={2012},
  publisher={International Society for Bayesian Analysis}
}

@inproceedings{yang2013poisson,
  title={On Poisson graphical models},
  author={Yang, Eunho and Ravikumar, Pradeep K and Allen, Genevera I and Liu, Zhandong},
  booktitle={Advances in Neural Information Processing Systems},
  pages={1718--1726},
  year={2013}
}

@article{chiquet2018variational,
  title={Variational inference for sparse network reconstruction from count data},
  author={Chiquet, Julien and Mariadassou, Mahendra and Robin, St{\'e}phane},
  journal={arXiv preprint arXiv:1806.03120},
  year={2018}
}

@article{hadiji2015poisson,
  title={Poisson dependency networks: Gradient boosted models for multivariate count data},
  author={Hadiji, Fabian and Molina, Alejandro and Natarajan, Sriraam and Kersting, Kristian},
  journal={Machine Learning},
  volume={100},
  number={2-3},
  pages={477--507},
  year={2015},
  publisher={Springer}
}

@article{inouye2016square,
  title={Square root graphical models: Multivariate generalizations of univariate exponential families that permit positive dependencies},
  author={Inouye, David I and Ravikumar, Pradeep and Dhillon, Inderjit S},
  journal={arXiv preprint arXiv:1603.03629},
  year={2016}
}

@article{inouye2016generalized,
  title={Generalized root models: beyond pairwise graphical models for univariate exponential families},
  author={Inouye, David I and Ravikumar, Pradeep and Dhillon, Inderjit S},
  journal={arXiv preprint arXiv:1606.00813},
  year={2016}
}

@article{inouye2017review,
  title={A review of multivariate distributions for count data derived from the Poisson distribution},
  author={Inouye, David I and Yang, Eunho and Allen, Genevera I and Ravikumar, Pradeep},
  journal={Wiley Interdisciplinary Reviews: Computational Statistics},
  volume={9},
  number={3},
  pages={e1398},
  year={2017},
  publisher={Wiley Online Library}
}

@article{yang2015graphical,
  title={Graphical models via univariate exponential family distributions},
  author={Yang, Eunho and Ravikumar, Pradeep and Allen, Genevera I and Liu, Zhandong},
  journal={The Journal of Machine Learning Research},
  volume={16},
  number={1},
  pages={3813--3847},
  year={2015},
  publisher={JMLR. org}
}
\end{document}